\documentclass[article, 10pt]{article}
\usepackage{amsmath}
\usepackage{amsthm}
\usepackage[dvips]{graphicx}
\usepackage{wrapfig}
\usepackage{pxfonts}
\usepackage{bm}

\newtheoremstyle{physics}{}{}{\itshape}{}{\bfseries}{}{\newline}{}

\theoremstyle{physics}

\newcommand\sgn{\operatorname{sgn}}
\newcommand\Ar{\mathcal{A}}

\begin{document}

\title{A note on cutting spin networks and \\ the area spectrum in loop quantum gravity}
\author{Yu Asato\thanks{email: yasato@ucdavis.edu} \\ {\it Department of Physics} \\ {\it University of California} \\ {\it Davis, CA 95616} \\ {\it USA}}
\date{June, 2015 \\ revised October, 2015}
\maketitle

\begin{abstract}
In this paper, I show that if a spin network is cut by a surface separating space-time into two regions, the sum of spins of edges crossing the surface must be an integer. This gives a restriction on the area spectrum of such surfaces, including black hole horizons, in loop quantum gravity.\\[10pt]
\end{abstract}

\section{Introduction}
A spin network is an embedded (possibly knotted) graph with its edges colored by integers or half-integers that label irreducible representations of SU(2), and with its vertices colored by intertwiners that are elements of invariant subspaces of  tensor product representations of irreducible representations of SU(2). Spin networks are essential in loop quantum gravity since they form a basis of the kinematical space of the theory \cite{Rov, RovSmo1, Baez}.

Even at a kinematical level, loop quantum gravity possesses intriguing predictions, one of which is the discretization of area. The spectrum of the area operator of a given surface, $\mathcal{S}$, is given by
\begin{equation} \label{eq3}
\Ar (\mathcal{S}) = 8\pi {l_{pl} }^2 \gamma \sum _i\sqrt{j_i(j_i+1)}~,
\end{equation}
where ${l_{pl}}$ is the Planck length, $\gamma $ is the Barbero-Immirzi parameter, and the sum is taken over all the edges intersecting the surface \cite{Rov, RovSmo2, AshLew}.

In the following, I first prove a proposition on the cutting of spin networks, and then show that the proposition gives a restriction on the area spectrum of a certain type of surfaces, which include horizons of black holes.

In most parts of the paper, for the sake of simplicity, I assume that cutting occurs on edges, rather than at vertices. However, the proof shown in the next section can be generalized into the case where cutting occurs at vertices. I will briefly describe it in section 3.

\section{Cutting of spin networks}

A sub-network is defined as a subgraph of a spin network with the same colorings of the corresponding edges and vertices. From the definition, we notice that a sub-network is not necessarily a spin network.

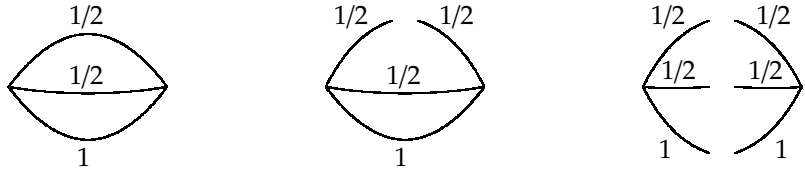
\begin{figure}[b] \centering 
\begin{picture}(300, 70)
\qbezier(0,35)(30,75)(60,35)
\qbezier(0,35)(30,-5)(60,35)
\qbezier(0,35)(30,30)(60,35)

\put(23,59){1/2}
\put(23,36){1/2}
\put(26,5){1}

\qbezier(120,35)(130,55)(145,60)
\qbezier(155,60)(170,55)(180,35)
\qbezier(120,35)(150,-5)(180,35)
\qbezier(120,35)(150,30)(180,35)

\put(123,59){1/2}
\put(163,59){1/2}
\put(143,36){1/2}
\put(146,5){1}

\qbezier(240,35)(250,55)(265,60)
\qbezier(275,60)(290,55)(300,35)
\qbezier(240,35)(250,34)(265,35)
\qbezier(275,35)(290,34)(300,35)
\qbezier(240,35)(250,15)(265,10)
\qbezier(275,10)(290,15)(300,35)

\put(243,59){1/2}
\put(283,59){1/2}
\put(247,38){1/2}
\put(280,38){1/2}
\put(246,8){1}
\put(290,8){1}
\end{picture}
\caption{An example of cutting a simple spin network. Left: the original spin network. Center: cutting of an edge; not a spin network nor a sub-network. Right: the cutting we are interested in; two sub-networks. We indeed cut two edges with $j=\frac{1}{2}$.}
\label{fig1}
\end{figure}

When we cut an edge of a spin network, the resulting graph may not be a sub-network of the original spin network. But we can always cut a spin network into a set of disconnected sub-networks (see Figure \ref{fig1}). On such a cutting, we find an interesting property:\\[10pt]
{\it \noindent
In order to cut a spin network into two sub-networks, one must cut an even number of edges colored by half-integers.\\[10pt]}
While this result is known as a ``folk theorem'' among some researchers in loop quantum gravity \cite{Rov2}, and an equivalent statement can be found in \cite{AshLew, AshLew2}, there seems to be no proof in the literature.

To prove this, we start by defining a useful function on a set of edges. Let $E$ be a set of colored edges. The signature of $E$ is defined as
\begin{equation}
\sgn (E):= (-1)^{n_F(E)}~, \
\end{equation}
where $n_F(E)$ is the number of edges in $E$ colored by half-integers. The signature of $E$ thus tells us whether $n_F (E)$ is odd or even. We further define the signature of a vertex, $v$, as
\begin{equation}
\sgn (v) := \sgn (E_v)\cdot \sgn (E_v^{loop})~,
\end{equation}
where $E_v$ is a set of edges connected to $v$, and $E_v^{loop}( \subset E_v)$ is a set of loops whose basepoints are $v$. The multiplication of $\sgn (E_v^{loop})$ is because here we would like to count the number of edges with half-integer spins coming into and going out of $v$, and hence loops contribute twice as much as non-loops do. We now have the following lemma. \\[10pt]
{\it \noindent 
Let $V_s$ be a set of all vertices of an arbitrary spin network, $s$. For arbitrary $v\in V_s$,
\begin{equation} \label{eq4-2}
\sgn (v) =1~.
\end{equation}
\begin{proof}
A tensor product representation of an odd number of half-integer irreducible representations of SU(2) cannot have an invariant subspace. Therefore, an intertwiner connecting an odd number of edges with half-integers never exists.
\end{proof}
}

Next, we consider a set of vertices, $V$. Defining $E_V:= \bigcup _{v\in V} E_v$, we call edges in $E_V$ connecting a pair of vertices in $V$ internal, and otherwise external. Let $E_V^{int}$ and $E_V ^{ext}$ denote the set of internal and external edges of $V$. Then, we define the signature of $V$ as
\begin{equation}
\sgn (V):= \sgn (E_V^{ext})~.
\end{equation}
Also, for $v\in V$, we define the sets of internal and external edges as
\begin{equation}
E_v^{int} := E_v \cap E_V^{int}~, \quad E_v^{ext} := E_v \cap E_V^{ext}~.
\end{equation}
That is, $E_v^{int}$ and $E_v^{ext}$ are sets of internal and external edges which are connected to $v$. We then have \\[10pt]
{\it \noindent
For a given spin network, s, and for arbitrary $V\subset V_s$,
\begin{equation} \label{eq2}
\sgn (V) =1~.
\end{equation}

\begin{proof}
Since $E_V=E_V^{int}\bigsqcup E_V^{ext}$, and $E_V= \bigcup _{v\in V} E_v$, we find $E_v=E_v^{int}\bigsqcup E_v^{ext}$, and thus
\begin{equation} \label{eq1}
n_F(E_v)=n_F(E_v^{int})+n_F(E_v^{ext})~.
\end{equation}
From the previous lemma, using eq.~(\ref{eq1}), we obtain, for arbitrary $v \in V$,
\begin{equation}
1=\sgn(v) = (-1)^{n_F (E_v) +n_F (E_v^{loop})} = (-1)^{n_F (E_v^{int})+n_F (E_v^{loop})+n_F(E_v^{ext})}~.
\end{equation}
But also
\begin{gather}
\sum _{v\in V} (n_F (E_v^{int}) + n_F (E_v^{loop})) =2n_F(E_V^{int})~, \\
\sum _{v\in V} n_F(E_v^{ext})=n_F(E_V^{ext})
\end{gather}
since the internal edges are counted twice, while the external edges once.
Hence, we find
\begin{eqnarray}
1&=&\prod _{v\in V}\sgn (v)=(-1)^{\sum _{v\in V} (n_F (E_v^{int})+n_F (E_v^{loop}))+\sum _{v\in V} n_F(E_v^{ext})} \nonumber \\
&=&(-1)^{2n_F (E_V^{int})+n_F(E_V^{ext})}=(-1)^{n_F(E_V^{ext})} \equiv  \sgn(V)~.
\end{eqnarray}
\end{proof}
}

Note that proving eq.~(\ref{eq2}) is indeed not necessary for the simple spin network in Figure \ref{fig1} since all of the edges on one side connect to a single intertwiner, and eq.~(\ref{eq4-2}), or equivalently the existence of the intertwiner, is sufficient to confirm the folk theorem. But, for general spin networks, the result is not so trivial, since the cut edges on each side of the surface may connect to many independent intertwiners, and we need to establish eq.~(\ref{eq2}).

Finally, we can easily prove the main claim. Suppose a spin network, $s$, is cut into sub-networks, $a$ and $b$. We notice that $E_{V_a}^{ext}(=E_{V_b}^{ext})$ is also a set of edges cut, $E_{cut}$. Then, eq.~(\ref{eq2}) tells that
\begin{equation} \label{eq13}
1=\sgn (V_a) = \sgn (E_{V_a}^{ext}) =\sgn (E_{cut})~,
\end{equation}
which is exactly the statement we wanted to prove.

It is also not at all difficult to generalize the statement to cutting of spin networks into many sub-networks, not just into two sub-networks (see the proof of (i) in the next section).

I also remark that the condition (\ref{eq13}) is sufficient for the existence of a spin network with a corresponding area spectrum. That is,\\[10pt]
{\it \noindent
For a given surface, $\mathcal{S}$, an arbitrary number of integers, $(l_i)_{i=1, ..., N}$, and an arbitrary even number of half-integers, $(l_i)_{i=N+1, ...,N+2M}$, there exists a spin network with an area $\Ar (\mathcal{S})=8\pi {l_{pl}}^2\gamma \sum _{i=1} ^{N+2M}\sqrt{l_i(l_i+1)}$.\\[0pt]}

We first prepare two vertices, $v_1$ and $v_2$, and $N+2M$ edges colored by $(l_i)_{i=1, ..., N+2M}$ which intersect $\mathcal{S}$ once and only once. Each edge has one end at $v_1$ and the other end at $v_2$.

Now we consider a tensor product representation, $l^{\otimes}:=\otimes _{i=1} ^{N+2M}l_i$. This representation can be decomposed into a sum of irreducible components. We then notice that those components are all integer representations since $(l_i)_{i=1, ..., N+2M}$ has an even number of half-integer representations.

\begin{figure}[t] \centering 
\begin{picture}(300, 75)

\qbezier(90,5)(110,85)(200,65)
\qbezier(90,5)(120,50)(200,65)
\qbezier(90,5)(170,-3)(200,65)

\put(145,73){\makebox(0,0){$l_1$}}
\put(154,59){\makebox(0,0){$l_2$}}
\multiput(162,47)(5,-6){3}{\circle*{2}}
\put(190,25){\makebox(0,0){$l_{N+2M}$}}

\put(80,5){\circle{20}}
\put(210,65){\circle{20}}

\put(65,5){\makebox(0,0){$\frac{n}{2}$}}
\put(225,65){\makebox(0,0){$\frac{n}{2}$}}

\put(95,0){\makebox(0,0){$v_1$}}
\put(196,72){\makebox(0,0){$v_2$}}

\put(95,55){\line(5,4){30}}
\put(95,55){\line(2,-3){38}}

\put(105,56){\makebox(0,0){$\mathcal{S}$}}

\end{picture}
\caption{The resulting spin network.}
\label{fig2}
\end{figure}
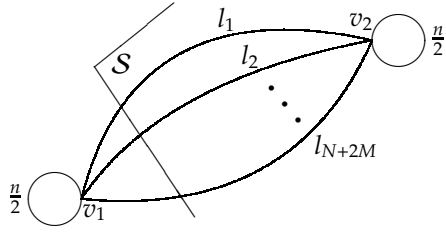

Let $n$ $(\in \mathbb{N})$ denote the highest-weight component in the decomposition. Introducing two additional edges both colored by $\frac{n}{2}$, we can obtain well-defined intertwiners at $v_1$ and $v_2$ by making the edge loop at each vertex because the irreducible decomposition of $l^{\otimes}\otimes \frac{n}{2} \otimes \frac{n}{2}$ contains an invariant subspace by its construction. This completes the construction of a desired spin network which is depicted in Figure \ref{fig2}.

\section{The minimal area of $\mathcal{S}$}

Let $\mathcal{S}$ be an arbitrary surface which separates space-time into two regions, such as the horizon of a black hole.
What makes the main result of the previous section physically interesting is that it brings the following two consequences:\\[10pt]
{\it \noindent
(i) $\mathcal{S}$ must be pierced by an even number of edges colored by half-integers. \\
(ii) The minimal area of $\mathcal{S}$ is given by a single intersection of an edge with $j=1$.

\begin{proof}[Proof of (i)]
Such a surface can cut spin networks into many sub-networks, but
\begin{equation}
E_{cut} =\bigsqcup _i E_{cut}^{(i)}~,
\end{equation}
where each of $E_{cut}^{(i)}$ follows the property given in section 2. Thus, $n_F (E_{cut})=\sum _i n_F(E_{cut}^{(i)})$, and 
\begin{equation}
1=\prod _i \sgn (E_{cut}^{(i)}) =\sgn (\bigsqcup _i E_{cut}^{(i)})=\sgn (E_{cut})~.
\end{equation}
\end{proof}

\begin{proof}[Proof of (ii)]
As stated previously, the area of $\mathcal{S}$ is given by eq.~(\ref{eq3}).
Now that it is forbidden to have edges colored by half-integers intersecting $\mathcal{S}$ singly, the minimal area involving half-integers is obtained from double piercing of edges with $j=\frac{1}{2}$:
\begin{equation}
\Ar _{\frac{1}{2}}^{min}(\mathcal{S})= 8\pi {l_{pl} }^2 \gamma \cdot 2\sqrt{\frac{1}{2}(\frac{1}{2}+1)}=8\sqrt{3}\pi {l_{pl} }^2 \gamma ~,
\end{equation}
while the minimal area involving integers is found to be
\begin{equation}
\Ar _{1}^{min}(\mathcal{S})=8\pi {l_{pl} }^2 \gamma \cdot \sqrt{1(1+1)}=8\sqrt{2}\pi {l_{pl} }^2 \gamma
\end{equation}
since edges colored by integers can still pierce $\mathcal{S}$ singly.
\end{proof}}

Also, we can easily generalize the result into the case where cutting occurs at vertices. First, if $\mathcal{S}$ cuts a spin network at vertices and on edges, we insert a bivalent vertex into each of the edges so that the cutting happens only at vertices. 
Second, if there are edges which have both ends on $\mathcal{S}$, but do not lie in $\mathcal{S}$, we also insert a bivalent vertex into each of the edges for the later convenience.
Now, vertices are naturally divided into three groups by $\mathcal{S}$: those in region 1, $V_1$, in region 2, $V_2$, and in $\mathcal{S}$. Let $E_{i}^{\mathcal{S}}$ denote a set of edges in region $i$ coming out of and going into $\mathcal{S}$. The number of edges with half-integer spins in $E_{i}^{\mathcal{S}}$ turns out to be even:
\begin{equation}
\sgn (E_{i}^{\mathcal{S}})=\sgn(E_{V_i} ^{ext})=1~.
\end{equation}
That is, the sum of spins of edges in each region which connect to $\mathcal{S}$ must be an integer. In the similar manner, we can also prove the statement about a surface without boundary which does not divide space-time into two regions, such as a 2-torus in a 3-torus.

Here, I did not change the gauge group from SU(2) to SO(3), nor require any assumptions on dynamics or supersymmetry to derive the result as is suggested in \cite{Dre, Cor, LinZha}. Rather, spin networks know by definition.

While this result probably has little effect on the area spectrum of large black holes, it changes the spectrum of small black holes considerably, and the consequences on black hole entropy have been investigated; details can be found in \cite{AguBar, BarLew}.

\section*{Acknowledgments}
I would like to thank Steven Carlip for his valuable suggestions. I would also like to express my thanks to Adam Getchell, Gabe Herczeg and Joe Mitchell for their helpful comments.

\end{document}